\documentclass{emulateapj}
\usepackage{apjfonts}

\shorttitle{Parameters of \object{LS~2883} and \object{PSR~B1259$-$63}}
\shortauthors{Negueruela et al.}

\begin{document}

%% LaTeX will automatically break titles if they run longer than
%% one line. However, you may use \\ to force a line break if
%% you desire.

\title{Astrophysical parameters of \object{LS~2883} and implications for the
\object{PSR~B1259$-$63} gamma-ray binary\altaffilmark{*}}
\altaffiltext{*}{Partially based on observations collected at the European
Southern Observatory, Paranal, Chile (ESO 282.D-5081) and the South African
Astronomical Observatory.}

%% Use \author, \affil, and the \and command to format
%% author and affiliation information.
%% Note that \email has replaced the old \authoremail command
%% from AASTeX v4.0. You can use \email to mark an email address
%% anywhere in the paper, not just in the front matter.
%% As in the title, use \\ to force line breaks.

\author{Ignacio Negueruela\altaffilmark{1}, Marc Rib\'o\altaffilmark{2},
Artemio Herrero\altaffilmark{3,4}, Javier Lorenzo\altaffilmark{1}, Dmitry
Khangulyan\altaffilmark{5}, Felix A. Aharonian\altaffilmark{6,7}\vspace{5mm}}

%\author{Ignacio Negueruela}
\affil{$^{1}$Departamento de F\'{i}sica, Ingenier\'{i}a de Sistemas y
Teor\'{i}a de la Se\~{n}al, Universidad de Alicante, Apdo. 99, E-03080
Alicante, Spain\email{ignacio.negueruela@ua.es}\email{javihd64315@gmail.com}}

%\author{Marc Rib\'o}
\affil{$^{2}$Departament d'Astronomia i Meteorologia, Institut de Ci\`ences del
Cosmos (ICC), Universitat de Barcelona (IEEC-UB), Mart\'{\i} i Franqu\`es 1,
E-08028 Barcelona, Spain\email{mribo@am.ub.es}}

%\author{Artemio Herrero}
\affil{$^{3}$Instituto de Astrof\'{i}sica de Canarias, E-38200 La Laguna,
Spain\email{ahd@iac.es}}
\affil{$^{4}$Departamento de Astrof\'{\i}sica, Universidad de La Laguna, Avda.
Astrof\'{\i}sico Francisco S\'anchez, s/n, E-38205 La Laguna, Spain}

%\author{Javier Lorenzo}
%\affil{$^{1}$Departamento de F\'{i}sica, Ingenier\'{i}a de Sistemas y Teor\'{i}a de
%la Se\~{n}al, Universidad de Alicante, Apdo. 99, E-03080 Alicante, Spain
%\email{javihd64315@gmail.com}}

%\author{Dmitry Khangulyan}
\affil{$^{5}$Institute of Space and Astronautical Science/JAXA, 3-1-1
Yoshinodai, Chuo-ku, Sagamihara, Kanagawa 252-5210,
Japan\email{khangul@astro.isas.jaxa.jp}}

%\author{Felix A. Aharonian}
\affil{$^{6}$Dublin Institute for Advanced Studies, 31 Fitzwilliam Place,
Dublin 2, Ireland\email{felix.aharonian@dias.ie}}
\affil{$^{7}$Max-Planck-Institut f\"ur Kernphysik, Saupfercheckweg 1, D-69117
Heidelberg, Germany}

%% Mark off your abstract in the ``abstract'' environment. In the manuscript
%% style, abstract will output a Received/Accepted line after the
%% title and affiliation information. No date will appear since the author
%% does not have this information. The dates will be filled in by the
%% editorial office after submission.

\begin{abstract}
Only a few binary systems with compact objects display TeV emission. The
physical properties of the companion stars represent basic input to understand the physical mechanisms behind the particle acceleration,
emission, and absorption processes in these so-called gamma-ray binaries. Here
we present high-resolution and high signal-to-noise optical spectra of
\object{LS~2883}, the Be star forming a gamma-ray binary with the
young non-accreting pulsar \object{PSR~B1259$-$63}, showing it to rotate faster
and be significantly earlier and more luminous than previously thought. Analysis of the interstellar lines
suggest that the system is located at the same distance as (and
thus is likely a member of) \object{Cen~OB1}. Taking the distance to the
association, $d=2.3$~kpc, and a color excess of $E(B-V)=0.85$ for
\object{LS~2883}, results in $M_{V}\approx-4.4$. Because of fast rotation, LS~2883 is oblate ($R_{{\rm eq}}\simeq9.7\:R_{\sun}$ and $R_{{\rm pole}}\simeq8.1\:R_{\sun}$) and presents a temperature gradient ($T_{{\rm eq}}\approx27\,500$~K, $\log g_{{\rm eq}}=3.7$; $T_{{\rm pole}}\approx34\,000$~K, $\log g_{{\rm pole}}=4.1$). If the star did not rotate, it
would have parameters corresponding to a late O-type star. We estimate its luminosity at $\log(L_{*}/L_{\sun})\simeq4.79$, and its mass at $M_{*}\approx30\:M_{\sun}$. The mass function
then implies an inclination of the binary system $i_{{\rm orb}}\approx23\degr$,
slightly smaller than previous estimates. We discuss the implications of these
new astrophysical parameters of \object{LS~2883} for the production of high
energy and very high energy gamma rays in the
\object{PSR~B1259$-$63}/\object{LS~2883} gamma-ray binary system. In
particular, the stellar properties are very important for prediction of the line-like bulk Comptonization component from the unshocked
ultra-relativistic pulsar wind.
\end{abstract}

%% Keywords should appear after the \end{abstract} command. The uncommented
%% example has been keyed in ApJ style. See the instructions to authors
%% for the journal to which you are submitting your paper to determine
%% what keyword punctuation is appropriate.


%% From the front matter, we move on to the body of the paper.
%% In the first two sections, notice the use of the natbib \citep
%% and \citet commands to identify citations.  The citations are
%% tied to the reference list via symbolic KEYs. The KEY corresponds
%% to the KEY in the \bibitem in the reference list below. We have
%% chosen the first three characters of the first author's name plus
%% the last two numeral of the year of publication as our KEY for
%% each reference.

\section{Introduction} \label{introduction}

Three binary systems containing a massive star and a compact object that
clearly display TeV emission are currently known: \object{LS~5039},
\object{LS~I~+61~303} and \object{PSR~B1259$-$63} (see, e.g.,
\citealt{ribo08r}). In \object{PSR~B1259$-$63} the compact object is a neutron
star, first discovered as a 47.7~ms radio pulsar, in a very wide and eccentric
orbit ($e=0.87$, $P_{{\rm orb}}= 1236.9$~d) around the Be star
\object{LS~2883}\footnote{Star 2883 in the catalog of Luminous Stars in the
Southern Milky Way \citep{stephenson71} is also known as
\object{CPD~$-63\degr$2495}. The use of SS~2883 should be avoided, as this
acronym refers to the catalog of emission-line stars by \citet{stephenson77},
which contains only 455 stars.} \citep{johnston92,johnston94}. A double-peaked
non-thermal and unpulsed radio outburst takes place around each periastron
passage \citep{johnston05}. This radio emission has recently been resolved at
AU scales by \cite{moldon11}, who place it outside the binary system.
Double-peaked outbursts have also been detected in soft and hard X rays, and at
TeV energies by HESS \citep{uchiyama09,grove95,aharonian05,aharonian09}. Upper
limits were obtained at GeV energies by EGRET around the 1994 periastron
passage \citep{tavani96}.

%Our understanding of the multi-wavelength emission is as follows
In the currently preferred scenario, the cold
ultra-relativistic wind of the pulsar interacts with the massive star wind
within the binary system. Relativistic electrons are accelerated in a shock
region where the pressures of both winds balance. These electrons produce
synchrotron radiation and upscatter UV-optical photons
from the companion star via Inverse Compton (IC), giving rise to broadband emission up to TeV gamma rays
(e.g., \citealt{kirk99,dubus06,khangulyan07,aharonian09}). GeV and TeV emission
can also be produced from IC scattering of stellar photons by electrons in the unshocked pulsar wind, depending on its Lorentz factor. Radio emission outside the binary system is produced by long-lived particles traveling away in a kind of cometary tail.

In these models, the companion star is the source of seed photons for IC scattering, expected to be the dominant radiation mechanism. Fundamental parameters, like
the luminosity or the effective temperature, have to be known accurately to
properly model, and thus be able to understand, how this gamma-ray binary
produces the observed multiwavelength properties (see \citealt{khangulyan07}).
\object{LS~2883} was observed in the optical by \citet{johnston94}, who
identified emission lines typical of a Be star. Based on the presence of strong
\ion{He}{1} emission, they concluded that it was earlier than B3. Arguing that
the star was likely located in the \object{Sagittarius Arm}, \citet{johnston94}
assumed a distance of 1.5~kpc and a spectral type B2e.
%, though
%warning that the typical luminosity of a B2\,Ve star would put it at only
%600~pc from the Sun \citep{johnston94}.

Here we report on new astrophysical parameters of \object{LS~2883}, a new
distance estimate and a new inclination of the binary orbit. These results will
allow for better modeling of the multiwavelength data acquired close to
the periastron passage of \object{PSR~B1259$-$63} in 2010 mid
December, the first covered by the {\it Fermi} and {\it AGILE} gamma-ray
missions.

\section{Observations and spectrum description} \label{observations}

Observations of \object{LS~2883} were obtained on 2009 August 28 using the
Ultraviolet and Visual Echelle Spectrograph (UVES; \citealt{dekker00}) mounted
on the 8.2~m Very Large Telescope (VLT) UT2 \emph{Kueyen} at Cerro Paranal,
Chile. UVES was used in dichroic 2 mode with cross-disperser CD\#2 in the blue
arm, giving coverage of the 373--499~nm range, and CD\#4 in the red arm,
allowing coverage of 565--946~nm with a small gap around 760~nm. We used Image
Slicer \#2 and a $0\farcs5$ slit, giving a resolving power $R\sim80\,000$.

In addition, intermediate-resolution spectra of \object{LS~2883} and three
nearby blue stars were obtained on 2006 May 8, using the unit spectrograph on
the 1.9-m telescope at the South African Astronomical Observatory (SAAO) in
Sutherland\footnote{{\tt
http://www.saao.ac.za/facilities/instrumentation/gratingspec/}} with grating
\#6 (600 lines~mm$^{-1}$). This configuration covers 380--560~nm with $R\approx2\,000$ (measured on arc frames).

\begin{figure}
%\epsscale{.80}
%\plotone{psr/f1.ps}
%\plotone{f1.ps}
\resizebox{\columnwidth}{!}{
\includegraphics[angle=0,clip]{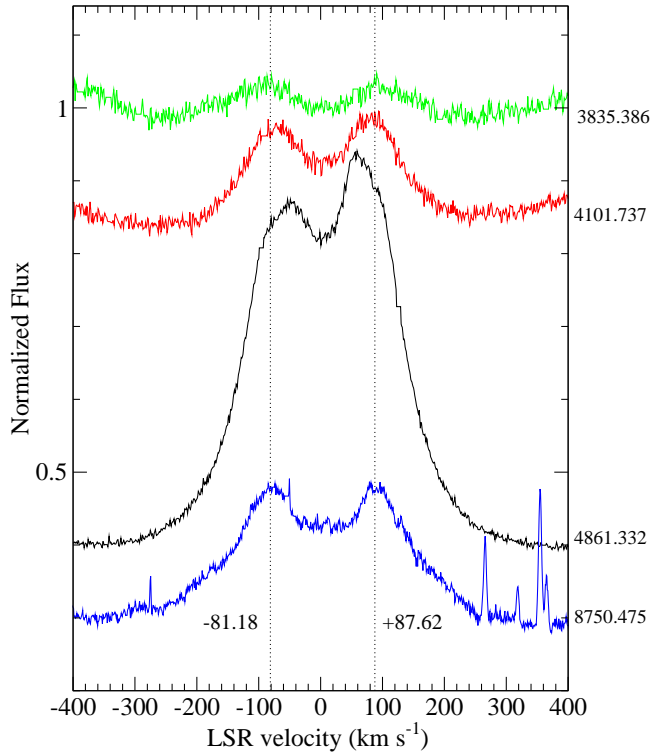}}
\caption{Representative emission lines in the VLT/UVES spectrum of LS~2883, in velocity space. Velocities are measured with respect to the Local Standard of Rest (LSR) assuming Solar motion
$+16.6\:{\rm km}\,{\rm s}^{-1}$ towards Galactic coordinates $\ell=53\degr$,
$b=+25\degr$. From top to bottom and vertically shifted for clarity, we show
H$\eta$ (green), H$\delta$ (red), H$\beta$ (black), and Pa~11 (blue). Their
central wavelengths are quoted in \AA\ to the right of the figure. The dotted
lines mark the position of the emission peaks in Pa~11, with velocities in
${\rm km}\,{\rm s}^{-1}$ annotated. 
%The upper Paschen lines have a very similar separation in velocity.
}
\label{fig:emission}
\end{figure}

 The spectrum of \object{LS~2883} is typical
of a Be star of early type with a well-developed disk \citep[cf.][]{johnston94}. Balmer lines have strong
double-peaked emission components, with the red peak clearly stronger than the
blue one (the two peaks are blended in H$\alpha$). Representative emission
lines are displayed in Figure~\ref{fig:emission}.
All Balmer lines higher than H$\gamma$ clearly show the photospheric absorption
wings, as the emission feature is definitely narrower than the absorption line, allowing the determination of effective gravity.
Paschen lines between Pa~11 and Pa~23 do not show any sign of the photospheric component. Other emission lines
typical of early-B stars, such as \ion{O}{1}~8448\AA\ or \ion{He}{1}~5875,
6678, 7065\AA\ appear as double-peaked features. The \ion{O}{1}~7775\AA\ triplet is blended into a strong single feature. There are many weak emission lines, mostly from \ion{Fe}{2} transitions, but also including \ion{Si}{2}~4128 \& 6347\AA. The
SAAO spectrum also shows strong emission lines from \ion{Fe}{2} 5019, 5169 \&
5316\AA. 
%(not covered in our VLT/UVES spectrum)
%, and other weaker features in
%the yellow region.

The equivalent width of H$\alpha$, EW(H$\alpha$), is $-54\pm2$\AA,
significantly higher than measured by \citet{johnston94}, $-40$\AA, though the
overall shape is very similar. The EW(H$\beta$) is $-4.5\pm0.4$ in the 2006
SAAO spectrum and $-4.6\pm0.2$ in the 2009 VLT/UVES spectrum, showing no
significant variability. As seen in Figure~\ref{fig:emission}, all the upper
Paschen and Balmer features have similar peak separation, with values
$170\pm10\:{\rm km}\,{\rm s}^{-1}$.

\section{Astrophysical parameters} \label{param}

\subsection{Distance and color excess estimates} \label{distance}

\begin{figure}
%\epsscale{.80}
%\plotone{psr/f3.ps}
%\plotone{f3.ps}
\resizebox{\columnwidth}{!}{
\includegraphics[angle=0,clip]{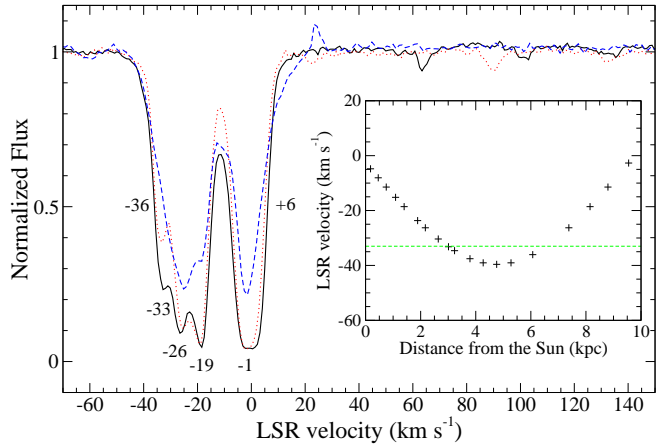}}
\caption{Interstellar lines in the VLT/UVES spectrum of LS~2883. The main panel
shows the components of three of the strongest interstellar lines:
\ion{Na}{1}~5890\AA\ (solid black), \ion{Na}{1}~5896\AA\ (dotted red) and
\ion{Ca}{2}~3934\AA\ (dashed blue) in velocity space. Peaks in the absorption and the extreme edges are annotated.
The inset shows the Galactic rotation curve (black crosses) along this line of
sight ($\ell=304\fdg2$, $b=-1\fdg0$) according to \citet{brand93}, using a
circular rotation velocity at the position of the Sun ($d_{{\rm GC}}=8.5$~kpc) of $220\:{\rm km}\,{\rm s}^{-1}$. The green dashed line indicates the velocity of the most negative peak measured towards LS~2883.}
\label{fig:abslines}
\end{figure}

Since model fitting can only give values for $T_{{\rm eff}}$ and $\log g$, we need an accurate distance to LS~2883 to derive $M_{*}$ and $L_{*}$. We use interstellar atomic lines in the spectrum of \object{LS~2883} to study
the radial velocity distribution of material along this line of sight
($\ell=304\fdg2$, $b=-1\fdg0$). 
%Absorbing clouds at different distances produce
%individual components of the interstellar feature, which sometimes appear
%separated. 
Figure~\ref{fig:abslines} displays three strong interstellar lines
in the spectrum of \object{LS~2883}, the two components of the \ion{Na}{1}
doublet and one of the members of the \ion{Ca}{2} doublet. 
%, while CH~4300\AA\ and CH$^{+}$~4232\AA\ have emission lines
%superimposed, distorting their shape. 
All three lines (and other interstellar lines, like CN~3875\AA) show the same morphology,
with two clearly separated components. The first component, with low positive
velocity, is likely due to nearby clouds associated with the \object{Southern
Coalsack}, which shows LSR velocities in the range $-10$ to $+8\:{\rm km}\,{\rm
s}^{-1}$ in CO maps \citep{nyman89}. The second feature is a combination of
several weaker components with negative velocities (between $-14\:{\rm
km}\,{\rm s}^{-1}$ and $-36\:{\rm km}\,{\rm s}^{-1}$), values produced by
clouds in the \object{Sagittarius Arm}. Very similar components are observed in the interstellar lines of \object{HD~112272} ($\ell=303\fdg5$, $b=-1\fdg5$), a B0.5\,Ia supergiant believed to be a member of \object{Cen~OB1}
\citep{hunter06}.

In this direction, the  \ion{H}{1} shell \object{GSH~305$+$01$-$24}, with dynamical $d=2.2\pm0.6$~kpc, has an average velocity
$-24\:{\rm km}\,{\rm s}^{-1}$, and is believed to be associated to
\object{Cen~OB1} \citep{mcclure02}. \object{LS~2883} shows interstellar components more negative than this value, suggesting that it is
beyond this shell. In fact, the Galactic rotation curve would place
\object{LS~2883} at $\ga$2.5~kpc (see inset of Figure~\ref{fig:abslines}). More distant stars along this sightline show narrow components at
$\approx-50\:{\rm km}\,{\rm s}^{-1}$, believed to arise from
clouds in the \object{Norma-Centaurus Arm}, at $d\ga4$~kpc \citep[and references therein]{kaper06}.
The lack of this component in \object{LS~2883} sets an upper limit to its
distance and firmly places it near \object{Cen~OB1}.

Photometric measurements of \object{LS~2883} in the literature \citep{klare77, schild83,westerlund89,drilling91}
suggest some variability, at the level of a few
hundredths of a magnitude, typical of Be stars. Taking $(B-V)=0.73$ as
representative, the intrinsic color corresponding to the model fit (see below),
$(B-V)_0=-0.28$, implies $E(B-V)=1.01$. Not all this reddening is interstellar,
as the disks of Be stars give rise to {\it circumstellar reddening} due to
stronger emission at longer wavelengths \citep[e.g.,][]{dachs88}. 
 The correlation between EW(H$\alpha$) and
$E^{\rm cs}(B-V)$ from \citet{dachs88}, valid solely for isolated Be stars,
predicts $E^{\rm cs}(B-V)=0.11$, higher than for most Be stars. Higher values are observed in Be/X-ray binaries, but values $\ga0.3$~mag are generally associated with transient events \citep[e.g.,][]{reig07}. Therefore it seems sensible to assume \object{LS~2883} to have interstellar $E(B-V)=0.8$--$0.9$~mag.

We have also estimated the color excess using four interstellar
diffuse bands (DIBs) with EW$>100$~m\AA\ in the spectrum of \object{LS~2883}: $\lambda$5780, $\lambda$5797 \citep{herbig93}, $\lambda$6202 and $\lambda$6614 \citep{rawlings03}. $\lambda$4430, is difficult
to measure. Though there is some dispersion, all support $E(B-V)$  between 0.8--0.9~mag. We shall thus assume $E(B-V)=0.85$.
Members of \object{Cen~OB1} near LS~2883 have $E(B-V)$ between 0.7 and 1.1 \citep{humphreys78}.
%The cataloged members
%closest in the sky to \object{LS~2883}, \object{HD~112272}, \object{HD~112366}
%and \object{HD~113432} have $E(B-V)=1.09$, $0.75$ and $1.02$,
%respectively\footnote{Slightly lower values are derived for \object{HD~112272}
%(1.00) and \object{HD~113432} (0.96) by \citet{winkler97}.}
%\citep{humphreys78}. 

The extinction law in this direction does not differ
significantly from the average $R=3.1$ law \citep{winkler97}. 
We observed two
luminous stars close to \object{LS~2883} from SAAO. For \object{LS~2888}
($7\arcmin$ away), we derive a spectral type B0.2\,III. Using photometry from
the literature \citep{klare77,schild83} and standard calibrations for intrinsic
colors \citep{fitzgerald70} and magnitudes \citep{turner80}, we estimate
$E(B-V)=0.83$ and $d=2.9\pm0.5$~kpc.
%, compatible within expected errors with the distance to \object{Cen~OB1}. 
For \object{LS~2882}, only
$30\arcsec$ away from \object{LS~2883}, we derive a spectral type B1\,IV.
Unfortunately, there is no accurate photometry published for this star. Its
2MASS $(J-K_{{\rm S}})$ color, however, indicates that it is reddened by an
amount very similar to \object{LS~2888}
%\footnote{For example, using the
%intrinsic colors of \citet{winkler97}, \object{LS~2888} has $E(J-K_{{\rm S}})=0.37$ and \object{LS~2882} has $E(J-K_{{\rm S}})=0.38$. If we assume that \object{LS~2883} has similar interstellar color excesses (and intrinsic colors from \citealt{martins06}), its circumstellar excesses take values $E^{{\rm
%cs}}(J-H)=0.22$ and $E^{{\rm cs}}(H-K)=0.38$. This ratio is typical of isolated
%Be stars, though the values are at the high end of those observed
%\citep{dougherty94}.}. 
The spectral types and magnitudes are fully compatible with the three stars being at similar distances.

Therefore direct measurements and indirect evidence are consistent with the
idea that \object{LS~2883} has interstellar $E(B-V)=0.85\pm0.05$ and is a member of \object{Cen~OB1}. The distance to this association has been studied by different authors, with most recent values converging towards $d=2.3\:{\rm kpc}$, though values up to $d=2.5\:{\rm kpc}$ are reported \citep[and references therein]{kaltcheva94,humphreys78}. We will assume that LS~2883 is located at the distance to the association, even though values as high as 2.8~kpc seem compatible with the data available. Values lower than 2.0~kpc seem to be excluded by the kinematic data contained in the interstellar lines to \object{LS~2883}.

\subsection{Spectrum modeling} \label{spectrum}

\begin{figure}
%\epsscale{.80}
\resizebox{\columnwidth}{!}{
\includegraphics[angle=0,clip]{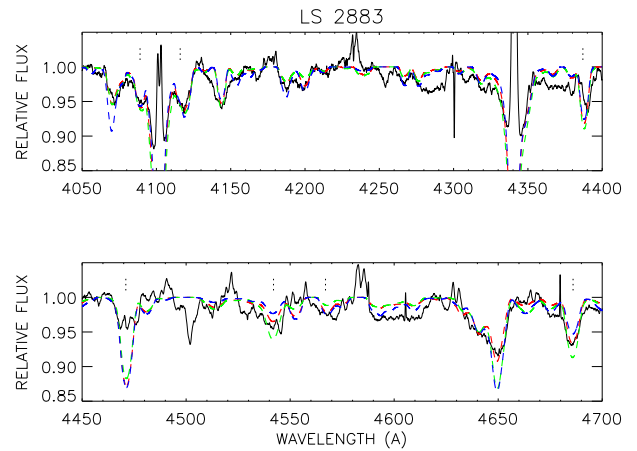}}
\caption{The rectified VLT/UVES blue spectrum of LS~2883 (solid black) and its
best model fit (dashed red). To illustrate the range of acceptable values, in the top panel we also show models calculated with $T_{{\rm eff}}=32\,000$~K and $\log g=3.60$ (blue dashes) and $4.10$ (green dashes). In the lower panel, we show models with $\log g=3.80$ and  $T_{{\rm eff}}=31\,000$~K (blue dashes) and  $T_{{\rm eff}}=34\,000$~K (green dashes). Features used for the analysis are marked with dashed lines. \label{fig:fit}}
\end{figure}

Stellar parameters have been calculated using \textsc{Fastwind}
\citep{puls05}, a spherical non-LTE model atmosphere code
with mass loss, developed to model detailed
spectral line profiles of hot and luminous stars. The best fit model is shown
in Figure~\ref{fig:fit}.

We determined the stellar rotational velocity using the Fourier transform method. We first degraded the spectrum to $R= 10\,000$ for a better SNR. 
Because of the large broadening, many lines are blended. We took unblended
line wings and mirrored them (thus producing a symmetric line) before calculating the Fourier transform. This procedure gives results consistent with measurements for the blended lines. Our values, which rest on the metallic
lines \ion{O}{2}~4254\AA\ and \ion{C}{2}~4267\AA, are $v_{{\rm rot}}\sin i=235\pm15\:{\rm km}\,{\rm s}^{-1}$ (projected rotational velocity; to be refined further below) and $\zeta= 115\pm 50\:{\rm km}\,{\rm s}^{-1}$ (so-called macroturbulence). Errors simply represent the dispersion of individual values. 
%Using the approximation by \citet{huang72}, the peak separation of the Paschen
%lines indicates that the minimum de-projected radius of the circumstellar disk
%is $7.6\:R_*$ (roughly 0.3~AU for the stellar radius estimates below).

\begin{table}
%\tabletypesize{\scriptsize}
\caption{Fitted and derived {\it apparent} stellar parameters for LS~2883
% and for the binary system
.\label{table:fit}}
\centering
\begin{tabular}{lc}
\tableline
\tableline
Parameter & Value \\
\tableline
$T_{{\rm eff}}$ (K)                     & $32\,000^{+2\,000}_{-1\,000}$\\ 
$\log g$                                & $3.80^{+0.30}_{-0.20}$\\
$R_{*}$\tablenotemark{1} ($R_{\sun}$)   & $9.0^{+1.8}_{-1.5}$\\
$\log(L_{*}/L_{\sun})$\tablenotemark{1} & $4.88^{+0.19}_{-0.17}$\\
$M_{*}$\tablenotemark{1} ($M_{\sun}$)   & $21.3^{+22.0}_{-9.0}$\\
$M_{*}$\tablenotemark{1,}\tablenotemark{2} ($M_{\sun}$)   & $26.6^{+23.5}_{-10.5}$\\
%$\log \dot M_{*}$ ($M_{\sun}$)&1.6$\times 10^{-7}$\\
%$i$\tablenotemark{1,}\tablenotemark{3} (\degr) & $35^{+0.0}_{-0.0}$\\
\tableline
\end{tabular}
\tablenotetext{1}{Assuming $d=2.3\pm0.4$~kpc and $E(B-V)=0.85\pm0.05$.}
\tablenotetext{2}{Deprojecting the rotational velocity with $i= 35\degr$, after applying the correction of \citet{fremat05}.}
\end{table}

Stellar parameters were obtained by visual fitting to the Balmer,
\ion{He}{1}/\ion{He}{2} and \ion{Si}{3}/\ion{Si}{4} lines \citep[see,
e.g.,][]{repolust04,simon10}. The fit is not as accurate as in these works,
since the heavy veiling by emission lines makes the continuum definition a
difficult task and some diagnostic lines either have emission components (such
as \ion{He}{1}~4471\AA) or are affected by nearby emission lines (as, for
example, \ion{Si}{3}~4552\AA). The best fit is obtained for $T_{{\rm
eff}}=32\,000^{+2\,000}_{-1\,000}\:{\rm K}$ and $\log g = 3.8^{+0.3}_{-0.2}$,
where the uncertainties have been obtained following the procedures discussed
in Sect.~6 of \cite{repolust04}.  A normal He abundance ($N$(He)/($N$(H)$+N$(He))= 0.09) is
consistent with the observed spectrum, although a slightly increased He abundance is possible. 
The lines typically used to estimate the mass-loss
rate are strongly contaminated by emission components, preventing us from
obtaining an accurate value. We have adopted the mass-loss rate expected for a spherical star of the 
luminosity derived below
according to the recipes of \cite{vink00}, resulting in a thin wind with 
no effect on the emergent profiles. 

Using $d=2.3\pm0.4$~kpc, $E(B-V)=0.85\pm0.05$ and $R=3.1\pm0.1$ we derive
$M_{V}=-4.4\pm0.4$. From the model fit spectrum and the absolute magnitude,  we
derive the values for $R_{*}$, $L_{*}$ and $M_{*}$ listed in
Table~\ref{table:fit}. The fitted
parameters correspond to an O9.5\,V star according to the observational scale
of \cite{martins05}, in good agreement with the morphological spectral classification that we derive from our spectra.

However, for fast rotators, there are strong differences in effective gravity between the polar and equatorial regions, resulting in important temperature gradients and oblateness \citep[and therein]{townsend04}. The binary mass function $f(M_{\rm NS})=1.53\:M_{\sun}$ \citep{johnston94} implies an orbital inclination
angle of $i_{{\rm orb}}=24\fdg7^{+5.9}_{-5.4}$ for a standard neutron star mass of $1.4\:M_{\sun}$.  \citet{melatos95} find a tilt of $10\degr$ between the orbital plane and circumstellar disk, expected to lie on the stellar equatorial plane. Therefore we are seeing LS~2883 under a low ($\sim35\degr$) inclination and the parameters obtained must mainly reflect the characteristics of the hotter polar region. 

 For a centrally condensed star, the critical rotational angular velocity (gravity and centrifugal forces are equal at the equator) is given by:

\begin{equation}
\Omega_{\rm crit}= \sqrt{\frac{8GM_{*}}{27 R^3_{\rm pole}}}\, .
\label{eqOmcrit}
\end{equation}

A star rotating at $\Omega_{\rm crit}$ will have $R_{{\rm eq}}=1.5\:R_{{\rm pole}}$ \citep{CO95}. In the case of LS~2883, with a deprojected $v_{{\rm rot}}=410\:{\rm km}\,{\rm s}^{-1}$, a rough estimation suggests that $R_{{\rm eq}}>1.1\:R_{{\rm pole}}$ and  $\omega \equiv \Omega_{*}/\Omega_{{\rm crit}} \ga 0.85$, meaning that 2D effects will be important.

For a start, in fast rotators, both the Full Width at Half Maximum (FWHM) of lines and the Fourier method underestimate $v_{{\rm rot}}\sin i$ \citep{townsend04,fremat05}.  We therefore correct $v_{{\rm rot}}\sin i$ using Fig.~8 of \citet{fremat05}, finding $v'_{{\rm rot}}\sin i=260\pm15\:{\rm km}\,{\rm s}^{-1}$. The deprojected velocity is then $v'_{{\rm rot}}\approx450\:{\rm km}\,{\rm s}^{-1}$, even closer to the critical value.

  Though a complete 2D treatment of the stellar surface is beyond our scope, we can perform a quick evaluation of the effects of fast rotation.
From the $T_{{\rm eff}}$ and $\log g$ derived from our analysis (apparent values), we estimate the $T_{{\rm eff}}$, $\log g$ and $M_{V}^0$ of a non-rotating star that would give the same apparent parameters when rotating at the same $v_{{\rm rot}}$ as LS~2883. For this, we use an iterative method based on the calculations of \cite{fremat05} and the tables of \citet{collins91}. With these new values, we calculate other stellar parameters and iterate again to take into account the change in the inclination due to the change in $M_{*}$. The results of this procedure are the stellar parameters that LS~2883 would have if it did not rotate (Table~\ref{table:param}, left panel) and its most likely actual parameters, taking into account fast rotation (Table~\ref{table:param}, right panel).

\begin{table}
%\tabletypesize{\scriptsize}
\caption{Estimated {\it actual} stellar parameters for LS~2883.\label{table:param}}
\centering
\begin{tabular}{lc|lc}
\tableline
\tableline
\multicolumn{2}{c|}{non-rotating}&\multicolumn{2}{c}{rotating} \\
\tableline
$T_{\rm eff}^0$ &33\,500~K &$i$&$33\degr$\\
$\log g^0$ & 4.0 &$\omega$& 0.88\\
$M^0_{V}$&-4.47& $T_{\rm eff}$~(eq) &27\,500~K\\
$R^0_{*}$ &$9.2\:R_{\sun}$ &$T_{\rm eff}$~(pole)&34\,000~K \\
$ \log (L^0_{*}/L_\odot)$ &4.98 & $\log g$~(eq) &3.7 \\
$M^0_{*}$ &$31\:M_{\sun}$&$\log g$~(pole) &4.1\\
&&$R_{\rm eq}$ &$9.7\:R_{\sun}$ \\
&&$R_{\rm pole}$& $8.1\:R_{\sun}$\\
&&$ \log (L_{*}/L_\odot)$& 4.79\\
\tableline
\end{tabular}
\end{table}

LS~2883 presents important differences in $T_{{\rm eff}}$ and $\log g$ between the polar and equatorial regions, rendering the apparent parameters little more than an approximate guess. The hypothetical non-rotating LS~2883 would have parameters roughly corresponding to an O8\,V star. Its observed spectrum is a consequence of fast rotation. The mass derived, $31\:M_{\sun}$, is somewhat high for these parameters, but is subject to large uncertainties, as it depends strongly on the corrections for fast rotation and the distance assumed.

\section{Discussion} \label{discussion}

Our spectroscopic observations and atmosphere model fitting have provided new
physical parameters for \object{LS~2883}, the massive star forming a gamma-ray
binary with the young non-accreting pulsar \object{PSR~B1259$-$63}. The higher
temperature and luminosity of the optical star presented
here,  $T_{\rm eq}\approx 27\,500$~K, $T_{\rm pole}\approx34\,000$~K,  
$L_*=2.3\times10^{38}$~erg~s$^{-1}$, as compared to previous
estimates, $T_{{\rm eff}}=23\,000$--$27\,000\:{\rm K}$,
$L_*=(0.3$--$2.2)\times10^{38}$~erg~s$^{-1}$ (see \citealt{khangulyan07}),
imply a significant revision of the parameters and conditions for the production
of nonthermal radiation in this binary system, especially in the gamma-ray
band. High (GeV) and very high (TeV) energy gamma-rays from this system can be
produced in two distinct regions: a) in the unshocked pulsar wind, i.e., a cold
ultra-relativistic outflow expanding outwards from the pulsar with Lorentz
factor $\Gamma \leq 10^6$; and (b) in the region of the terminated pulsar wind.
In both regions, the dominant gamma-radiation mechanism is IC scattering of
relativistic electrons. While the Very High Energy (VHE) emission detected by
HESS is most likely linked to the multi-TeV electrons accelerated after
termination of the wind, GeV gamma-ray emission can be effectively produced
also by the unshocked pulsar wind. The higher luminosity of the optical star
obviously implies an enhanced interaction rate and consequently higher
luminosity of IC gamma-rays ($L_\gamma \propto L_{\rm *}$). The {\it Fermi} and
{\it AGILE} gamma-ray missions should be able to detect, at epochs close to the
periastron, the line-type emission of the unshocked pulsar wind
\citep{kirk99,ball00,khangulyan07}, and in this way measure its Lorentz factor
and mechanical power.

At first glance, the enhanced luminosity of the optical star should have a 
weaker impact on the gamma-ray emission related to the termination of the 
pulsar wind, since gamma-ray production is expected to proceed in the 
saturation regime. However, the enhanced optical luminosity does affect the
gamma-ray emission for two reasons. Firstly, it introduces a significant
orbital dependence of the gamma-ray flux due to the Compton deceleration of the
unshocked pulsar wind. This effect has been realized by \citet{khangulyan07},
who indicated that the deficit of VHE emission observed by HESS close to the
periastron passage may be explained by this effect, provided that the
luminosity of the optical star is as large as
$\sim4\times10^{38}$~erg~s$^{-1}$. Although the value found  here is smaller by a factor $1.7$, the Compton drag effect should be combined with gamma-gamma absorption, since for the updated orbital
inclination the maximal attenuation (by a factor of $2$) occurs close to periastron passage. We note,
however, that given the large orbital separation the total absorbed energy will
be small, thus the contribution from the electromagnetic cascade should be
negligible.

Finally, the production rate of VHE gamma-rays is affected by the increase of
the stellar temperature and the change of the orbital inclination. The higher
photon temperature leads to a more pronounced impact of the Klein-Nishina
effect. This, together with the smaller orbital inclination obtained here,
should result in a weaker orbital phase dependence of the VHE gamma-ray
production, in agreement with recent HESS observations
\citep{aharonian09,kerschhaggl11}.

It is clear that the new physical parameters of \object{LS~2883} reported here
have to be taken into account in the interpretation of the high and very
high energy gamma-ray observations of the gamma-ray binary
\object{LS~2883}/\object{PSR~B1259$-$63}, such as the multiwavelength campaigns close to the 2010 December periastron
passage, which include for the first time the participation of the {\it
Fermi} and {\it AGILE} gamma-ray missions.

\acknowledgments

We thank Dr.~I.D.H.~Howarth and the referee, Dr.~J.~Puls,
for very valuable suggestions, and J.~Mold\'on for useful comments.
This research is partially supported by the Spanish MICINN (grants FPA2010-22056-C06-02, AYA2008-06166-C03-01/03, AYA2010-21697-C05-04/05 and CSD2006-70 and FEDER funds); and by the Generalitat
Valenciana (ACOMP/2009/164) and Gobierno de Canarias
(ProID2010119). M.R. acknowledges financial support from
MICINN and European Social Funds through a \emph{Ram\'on y Cajal} fellowship.

%% To help institutions obtain information on the effectiveness of their
%% telescopes, the AAS Journals has created a group of keywords for telescope
%% facilities. A common set of keywords will make these types of searches
%% significantly easier and more accurate. In addition, they will also be
%% useful in linking papers together which utilize the same telescopes
%% within the framework of the National Virtual Observatory.
%% See the AASTeX Web site at http://www.journals.uchicago.edu/AAS/AASTeX
%% for information on obtaining the facility keywords.

%% After the acknowledgments section, use the following syntax and the
%% \facility{} macro to list the keywords of facilities used in the research
%% for the paper.  Each keyword will be checked against the master list during
%% copy editing.  Individual instruments or configurations can be provided 
%% in parentheses, after the keyword, but they will not be verified.

{\it Facilities:} \facility{VLT:Kueyen (UVES)}

%% The reference list follows the main body and any appendices.
%% Use LaTeX's thebibliography environment to mark up your reference list.
%% Note \begin{thebibliography} is followed by an empty set of
%% curly braces.  If you forget this, LaTeX will generate the error
%% "Perhaps a missing \item?".
%%
%% thebibliography produces citations in the text using \bibitem-\cite
%% cross-referencing. Each reference is preceded by a
%% \bibitem command that defines in curly braces the KEY that corresponds
%% to the KEY in the \cite commands (see the first section above).
%% Make sure that you provide a unique KEY for every \bibitem or else the
%% paper will not LaTeX. The square brackets should contain
%% the citation text that LaTeX will insert in
%% place of the \cite commands.

\clearpage

\end{document}